\documentclass{article}

\newcommand{\be}{\begin{equation}}
\newcommand{\ee}{\end{equation}}
\newcommand{\bea}{\begin{eqnarray}}
\newcommand{\eea}{\end{eqnarray}}

\begin{document}

\title{ %
\begin{flushright}
\small DTP-MSU-0019 \\ \small \tt gr-qc/0101100
\end{flushright} \vspace{.5cm}
 Quantization Near Violent Singularities
                 In Einstein-Yang-Mills Black Holes
                 \footnote{Contribution to the 9th Marcel Grossmann meeting (MG9), Rome, July 2000
}}

\author{{\large D.V. Gal'tsov \footnote{Supported by RFBR grant 00-02-16306.}}\\
\it Department of Theoretical Physics, \\
\it Moscow State University, 119899, Moscow,
Russia\\
E-mail: galtsov@grg.phys.msu.su }

\maketitle

\begin{abstract}
Classical singularities inside black holes in the
Einstein-Yang-Mills theory exhibit unusual features. Only for
discrete values of the black hole mass one encounters
singularities of the Schwarzschild type (timelike) and the
Reissner-Nordstr\"om type (spacelike). For a generic mass the
approach to singularity is not smooth: the metric oscillates
with an infinitely growing amplitude and decreasing period. In
spite of some similarity with the BKL oscillations, here the
behavior is not chaotic. However the oscillation amplitude
exceeds classical limits after few cycles, so the question
arises how this  behavior gets modified by quantum effects. We
discuss this issue both in the framework of QFT and in the
string  theory.
\end{abstract}

From all known static black hole solutions in General
Relativity the Einstein-Yang-Mills (EYM) black holes are
distinguished by a very complicated singularity
structure~\cite{DoGaZo97,GaDoZo97,BrLaMa98} (for a review
see~\cite{VoGa99}). First, we used to think that, in a given
theory, the static spherically symmetric black holes are unique
like the Schwarzschild black hole in vacuum, the
Reissner-Nordstr\"om one in electrovacuum, the dilatonic black
hole in the Einstein-Maxwell-dilaton theory and so on. In the
EYM case not only one has an infinite family of exterior
solutions with the same mass (and zero charge), but, rather
independently of the exterior behavior, these black holes can
have qualitatively different internal structure depending on
the value of mass and the second discrete parameter, the number
of nodes of the YM function. Namely, for certain discrete
masses, the singularity can be of the Schwarzschild type
(timelike), for other discrete masses it turns out to be of the
Reissner-Nordstr\"om type (spacelike), while a generic
continuously varied mass corresponds to an oscillating
singularity of a new type.

The oscillation regime starts after the  mass function $m(r)$,
defined via $
ds^2=(1-2m(r)/r)\sigma^2(r)dt^2-dr^2/(1-2m(r)/r)-r^2d\Omega,$
reaches a local minimum such that $g_{rr}^{-1}$ is very small
('almost' event horizon). Then an exponentially fast jump of
$m(r)$ is observed (when moving downward to the singularity)
very similar the the mass-inflation near the internal Cauchy
horizons~\cite{BrLaMa98}. This pictures reproduces itself in
more and more violent cycles which can be described as follows.
 The $k$-th cycle
starts at the $k$-th local minimum of the mass function $m$ at
some $r_k$. Then, as $r$  goes  to some point $R_k<r_k$, $m$
grows exponentially fast till it stabilizes on a very large
value at $R_k$, while the second metric function $\sigma$ is
decreased by many orders of magnitude. In the region $r<R_k$
the mass function reaches a horizontal plateau
 $m\approx M_k\equiv m(R_k)$, and
 $\sigma\approx \sigma_k\equiv \sigma(R_k)$.
The plateau stretches as far as many orders of magnitude of $r$
towards the singularity, during which period the geometry is
approximately Schwarzschild with a very large mass $M_k$. Then,
at end of the plateau, $m$ catastrophically falls down reaching
a very deep minimum at some $r_{k+1}$, after which the next
oscillation cycle starts. The sequence  of ratios
$x_k=(r_k/R_k)^2\gg1$ by  satisfies the equation~\cite{GaDoZo97}
$x_{k+1}=e^{x_k}/x_{k}^3$. In terms of $x_k$ the ratio of the
neighboring oscillation periods  is ${r_{k+1}}/{r_k}=x_k\,
e^{-x_k/2},$ (note that  $r_k\gg r_{k+1}$).   $M_k$ increases as
$\frac{M_k}{M_{k-1}}=\frac{e^{x_k/2}}{x_k},$ while $\sigma_k$
decrease as $\frac{\sigma_{k+1}}{\sigma_{k}}=e^{-x_k/2}$. The
oscillations are associated with very large values of $w'$ at
some very tiny intervals, so that $w$ itself remains bounded
and still has a finite limit.

The interior of a static spherically symmetric black hole can be
presented as the Kantowski-Sachs anisotropic universe. So the
oscillations inside the EYM black hole can be equally
attributed to a cosmological model filled with the YM matter.
Oscillating approach to singularity in cosmology usually is
associated with the Belinski-Khalatnikov-Lifshitz (BKL)
solution, which, however, belongs to the Bianchi IX type. The
qualitative difference of our case is that the EYM oscillations
are not chaotic: they can be described by an autonomous
two-dimensional dynamical system.

Clearly, classical limits are violated after a few oscillation
cycles. Quantum theory is expected to smoothen such a behavior.
The first approach to show that this is indeed the case is
particle creation calculations in the Kantowski-Sachs cosmology.
Very similar to the earlier treatment of particle creation in
various  anisotropic models one can show that the solution is
modified to the form of the power-low mass-inflation $m\sim
c/r^\alpha$.

Another approach is related to the string theory, where the YM
field is associated with the open strings. The corresponding
effective action (as well as that for D-branes) is given by the
Born-Infeld type non-Abelian lagrangian which is endowed with
an intrinsic damping mechanism at the large field strength.
This lagrangian sums up all $\alpha'$ corrections, so, in spite
of the fact that it is obtained in the constant
 field approximation, one can hope that it provides a reasonable
model for probing the string effects near the EYM singularity
(gravity being still treated classically).

Actual calculations~\cite{DyGa00} show that the mass function
monotonously approach the singularity which is always of the
Schwarzchild type and  exhibits a power-low inflation behavior.
This result was obtained in collaboration with V.V.~Dyadichev.
So both QFT and string quantum effects 'regularize' violent
oscillations near the EYM generic singularity. It is worth
noting that no Cauchy horizons appears, so the strong cosmic
censorship conjecture holds inside the 'realistic' EYM black
holes.

\end{document}